\documentclass[10pt,letterpaper]{article}
\usepackage{opex3}
\usepackage{multicol}
\usepackage{multirow}
\usepackage{epstopdf}
\usepackage{cite}
\usepackage{graphicx}
\usepackage{slashbox}

\begin{document}

\title{Silicon avalanche photodiode operation and lifetime analysis for small satellites} 
\author{Yue Chuan Tan*, Rakhitha Chandrasekara,\\
Cliff Cheng, and Alexander Ling}

\address{Centre for Quantum Technologies, National University of Singapore, Block S15, 3 Science Drive 2, Singapore 117543}

\email{*cqttyc@nus.edu.sg} 



\begin{abstract}
Silicon avalanche photodiodes (APDs) are sensitive to operating temperature fluctuations and are also susceptible to radiation flux expected in satellite-based quantum experiments. We introduce a low power voltage adjusting mechanism to overcome the effects of in-orbit temperature fluctuations. We also present data on the performance of Si APDs after irradiation ($\gamma$-ray and proton beam).  Combined with an analysis of expected orbital irradiation, we propose that a Si APD in a 400 km equatorial orbit may operate beyond the lifetime of the satellite.
\end{abstract}



\bibliographystyle{osajnl}
\bibliography{ref}

\begin{thebibliography}{10}
\newcommand{\enquote}[1]{``#1''}

\bibitem{ursin07}
R.~Ursin, F.~Tiefenbacher, T.~Schmitt-Manderbach, H.~Weier, T.~Scheidl,
  M.~Lindenthal, B.~Blauensteiner, T.~Jennewein, J.~Perdigues, P.~Trojek,
  B.~ï¿½mer, M.~Fï¿½rst, M.~Meyenburg, J.~Rarity, Z.~Sodnik, C.~Barbieri,
  H.~Weinfurter, and A.~Zeilinger, \enquote{{Entanglement-based quantum
  communication over 144 km},} Nat. Phys. \textbf{3}, 481--486 (2007).

\bibitem{meyerscott11}
E.~Meyer-Scott, Z.~Yan, A.~MacDonald, J.-P. Bourgoin, H.~H\"ubel, and
  T.~Jennewein, \enquote{How to implement decoy-state quantum key distribution
  for a satellite uplink with 50-db channel loss,} Phys. Rev. A \textbf{84},
  062326 (2011).

\bibitem{ursin08}
R.~Ursin, \enquote{Space-quest,} in \enquote{IAC Microgravity Sciences and
  Processes Symposium,}  (2008).

\bibitem{prochazka09}
I.~Prochazka and Y.~Fumin, \enquote{Photon counting module for laser time
  transfer via earth orbiting satellite,} J. Mod. Opt. \textbf{56},
  253--260 (2010).

\bibitem{krainak10}
M.~A. Krainak, A.~W. Yu, Y.~Guangning, S.~X. Li, and X.~Sun,
  \enquote{Photon-counting detectors for space-based laser receivers,}
  Proc. SPIE Int. Soc. Opt. Eng. \textbf{7608}, 760827--1 (2012).

\bibitem{woellert11}
K.~Woellert, P.~Ehrenfreund, A.~J. Ricco, and H.~Hertzfeld, \enquote{Cubesats:
  Cost-effective science and technology platforms for emerging and developing
  nations,} Adv. Space Res. \textbf{47}, 663--684 (2011).

\bibitem{sun04}
X.~Sun, M.~A. Krainak, J.~B. Abshire, D.~Spinhirne, James, C.~Trottier,
  M.~Davies, H.~Dautet, G.~R. Allan, A.~T. Lukemire, and J.~C. Vandiver,
  \enquote{Space-qualified silicon avalance-photodiode single-photon-counting
  modules,} J. Mod. Opt. \textbf{51}, 1333--1350 (2004).

\bibitem{apdrad}
J.~Kataoka, T.~Toizumi, T.~Nakamori, Y.~Yatsu, Y.~Tsubuku, Y.~Kuramoto,
  T.~Enomoto, R.~Usui, N.~Kawai, H.~Ashida, K.~Omagari, K.~Fujihashi,
  S.~Inagawa, Y.~Miura, Y.~Konda, N.~Miyashita, S.~Matsunaga, Y.~Ishikawa,
  Y.~Matsunaga, and N.~Kawabata, \enquote{In-orbit performance of avalanche
  photodiode as radiation detector on board the picosatellite cute-1.7+apd ii,}
  J. Geophys. Res. \textbf{115}, A05204 (2010).

\bibitem{zappa96}
F.~Zappa, S.~Cova, M.~Ghioni, A.~Lacaita, and C.~Samori, \enquote{Avalanche
  photodiodes and quenching circuits for single-photon detection,} Appl.
  Opt. \textbf{35}, 1956--1976 (1996).

\bibitem{peloso08}
M.~P. Peloso, I.~Gerhardt, C.~Ho, A.~Lamas-Linares, and C.~Kurtsiefer,
  \enquote{Daylight operation of a free space, entanglement-based quantum key
  distribution system,} New J. Phys.\textbf{11}, 045007 (2009).

\bibitem{stipcevic10}
M.~Stip\v{c}evi\'{c}, H.~Skenderovi\'{c}, and D.~Gracin,
  \enquote{Characterization of a novel avalanche photodiode for single photon
  detection in vis-nir range,} Opt. Express \textbf{18}, 17448--17459 (2010).

\bibitem{morong12}
W.~Morong, D.~Oi, and A.~Ling, \enquote{Quantum optics for space platforms,}
  Opt Photonics News (October 2012, 43--49).

\bibitem{SROUR1988}
J.~R. Srour and J.~M. McGarrity, \enquote{Radiation effects on microelectronics
  in space,} Proc. IEEE \textbf{76}, 1443--1469 (1988).

\bibitem{Johnston2000}
A.~H. Johnston, \enquote{Radiation damage of electronic and optoelectronic
  devices in space,} in \enquote{4th International Workshop on Radiation
  Effects on Semiconductor Devices for Space Application,}  (2000).

\bibitem{sun97}
X.~Sun, D.~Reusser, H.~Dautet, and J.~Abshire, \enquote{Measurement of proton
  radiation damage to si avalanche photodiodes,} IEEE Trans. Electron
  Devices \textbf{44}, 2160--2166 (1997).

\bibitem{spenvis}
\emph{The Space Environment Information System. http://www.spenvis.oma.be/}.

\bibitem{Raymond1987}
J.~Raymond and E.~Petersen, \enquote{Comparison of neutron, proton and gamma
  ray effects in semiconductor devices,} IEEE Trans. Nucl. Sci.
  \textbf{34}, 1622--1628 (1987).

\bibitem{Akkerman2001}
A.~Akkerman, J.~Barak, M.~Chadwick, J.~Levinson, M.~Murat, and Y.~Lifshitz,
  \enquote{Updated NIEL calculations for estimating the damage induced by
  particles and gamma-rays in Si and GaAs,} Radiat. Phys. Chem.
  \textbf{62}, 301--310 (2001).

\bibitem{Pease1988}
R.~L. Pease, A.~H. Johnston, and J.~L. Azarewicz, \enquote{Radiation testing of
  semiconductor devices for space electronics,} Proc. IEEE
  \textbf{76}, 1510--1526 (1988).

\bibitem{dale92}
C.~J. Dale, R.~A. Howard, P.~W. Marshall, B.~Cummings, L.~Shamey, and A.~W.
  Delamere, \enquote{Spacecraft displacement damage dose calculations for
  shielded ccds,} Proc. SPIE Int. Soc. Opt. Eng. \textbf{1656}, 476 (1992).

\bibitem{Summers1993}
G.~P. Summers, E.~A. Burke, P.~Shapiro, and S.~R. Messenger, \enquote{Damage
  correlations in semiconductors exposed to gamma, electron and proton
  radiations,} IEEE Trans. Nucl. Sci. \textbf{40}, 1372--1379
  (1993).

\bibitem{Jun2003a}
I.~Jun, M.~A. Xapsos, S.~R. Messenger, E.~A. Burke, R.~J. Walters, G.~P.
  Summers, and T.~Jordan, \enquote{Proton nonionizing energy loss ({NIEL}) for
  device applications,} IEEE Trans. Nucl. Sci. \textbf{50},
  1924--1928 (2003).

\bibitem{subash11}
S.~Sachidananda and A.~Ling, \enquote{Bias voltage control of avalanche
  photo-diode using a window comparator,}  (IEEE Summer Topicals, 2011).

\end{thebibliography}


\section{Introduction}
\label{sec:intro}
Terrestrial free-space optical quantum communication is currently limited in range by the need for line-of-sight locations \cite{ursin07}. To extend the range to the continental scale and beyond, it has been proposed that orbiting satellites carrying quantum communication systems can either transmit or receive single photons from widely separated ground stations \cite{meyerscott11,ursin08}. In either scenario single photon detectors operating in Geiger mode must be carried by satellites for verifying the onboard photon source, or for detecting the received photons. Silicon avalanche photodiodes (APDs) for single photon detection in the range from 400 nm to 1000 nm have been previously demonstrated in space\cite{prochazka09,krainak10} onboard conventional spacecraft.

Conventional satellites, however, are costly and take years to build. Cost effective space missions to Low Earth Orbit (LEO) can be rapidly developed using smaller spacecraft such as nanosatellites that have a mass below 10 kg. A widely accepted nanosatellite standard is the cubesat platform designed around commercial-off-the-shelf (COTS) technologies \cite{woellert11}. 

The cubesat footprint is a 10 cm cube (1U) with a mass below one kilogram and limited power available (typically not exceeding 1.5 W). Apart from the essential sub-systems (radio, battery and flight controller) about 0.3U of volume can be set aside for a payload. Cubesats can be stacked to form larger systems. Launching single photon experiments into LEO via cubesats could open up opportunities for demonstrating long distance Bell measurement and intercontinental quantum key distribution at comparatively low cost. The drawback is that cubesat payloads must tolerate space radiation \cite{sun04} and temperature fluctuations (between 0$^\circ$C and 20$^\circ$C every 100 minutes) \cite{apdrad}. This is mitigated on larger satellites, but is a challenge for cubesats where heating power and shielding are limited. Reduced radiation tolerance of COTS components is balanced against a shorter orbital lifetime (typically below 1 year) as cubesats are often launched into inner LEO trajectories (200-400 km).

Motivated by the challenges of operating quantum communication experiments on 1U cubesats we have designed and implemented a low power APD control circuit.  We have also exposed APDs to radiation and observed how they can continue to be operated with our novel control circuit.  
In this paper we demonstrate our approach to combat in-orbit temperature fluctuations and the results from our radiation exposure tests.
We conclude with a discussion about possible development paths for Geiger mode Si APDs in nanosatellites.

\section{Low power operation of APDs in a dynamic thermal environment}
\label{sec:wincom}

\begin{figure}  
		\includegraphics[width=\textwidth]{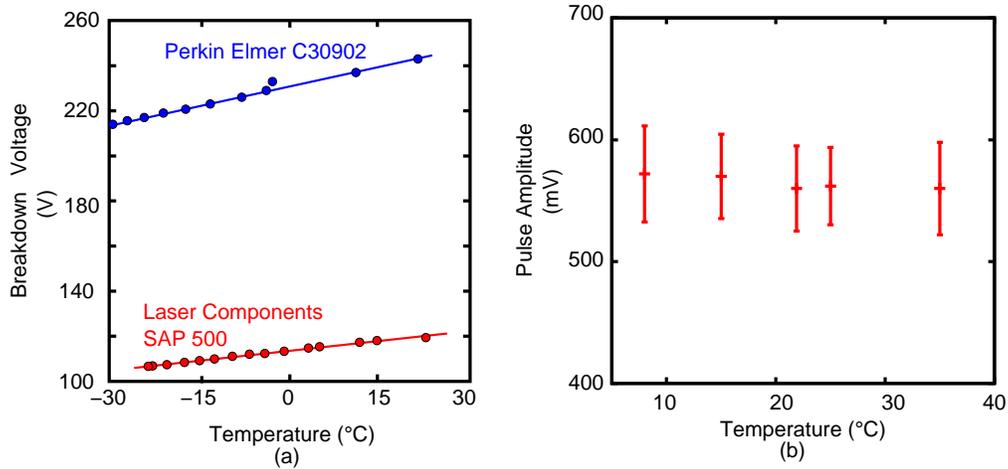} 
		\caption{(a) Breakdown voltage against operating temperature for two devices.  (b) SAP500 pulse amplitude at different temperature using the window comparator mechanism. The error bars are the Full Width at Half-Maximum of the amplitude distribution. (Color online).} \label{fig:tempapd} \end{figure}

Geiger mode Si APDs are operated with a bias voltage (V$_{\mathrm{b}}$) greater than the device breakdown voltage (V$_{\mathrm{br}}$). 
An incident photon absorped by the APD triggers an electron avalanche that is detected externally. 
Any avalanche must be stopped to prepare the APD for the next event by an active or passive quenching circuit \cite{zappa96}.  
The detection efficiency increases with the excess voltage, V$_{\mathrm{E}}$=V$_{\mathrm{b}}$ - V$_{\mathrm{br}}$, since a larger V$_{\mathrm{E}}$ increases the probability of triggering an avalanche.
A fixed detection efficiency requires a fixed V$_{\mathrm{E}}$.
Typically APDs are operated at a fixed low temperature to minimize dark counts (spontaneous avalanche due to thermal excitation) and to maintain a constant APD breakdown voltage (see Fig.~\ref{fig:tempapd}(a)) enabling constant detection efficiency for a fixed bias voltage. 

Operating APDs on a cubesat is challenging due to the limited power available.  
A COTS module running a single APD can consume over 2W of electrical power with a large fraction dedicated to maintaining a fixed sub-zero temperature, exceeding the cubesat power budget.  
However, cooling is not always necessary when APDs are operated in coincidence detection, e.g. when detecting correlated photon-pairs.
The typical dark count rate at room temperature is in the range of 20-50 kcps.  In comparison, quantum key distribution has been successfully demonstrated even with a background count rate of 250 kcps \cite{peloso08}.
Operating at fixed temperature is also not necessary if individual APDs are calibrated for the range of possible operating temperatures.
However, this can be tedious and may be inaccurate after the APDs have accumulated radiation damage.

The other consideration is the APD breakdown voltage.  On cubesats, it is necessary to convert battery voltage (between 5 and 12 V) to V$_{\mathrm{b}}$ values that exceed 100 V.  
In Fig.~\ref{fig:tempapd}(a), we present the measured breakdown voltages for two different devices: the Perkin-Elmer C30902 and the Laser Components SAP500. The SAP500 operates about 100V below the C30902 requiring less operating power and with no observed loss in detection efficiency \cite{stipcevic10} hence, development efforts have been focused around the SAP500.

                \begin{figure} \centering \includegraphics[width=0.7\linewidth,clip]{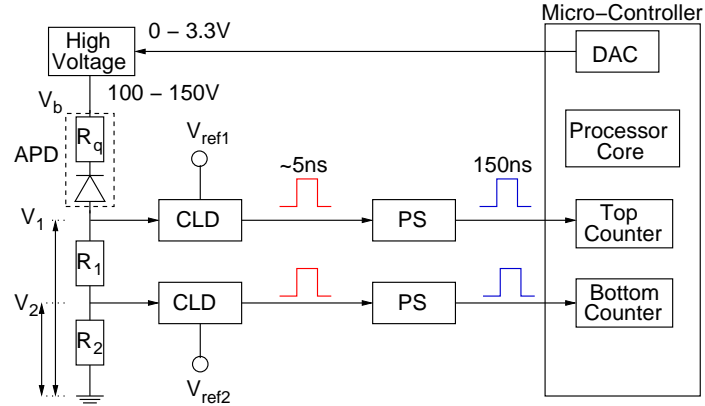} \caption{Setup of window comparator mechanism. 
Bias voltage (V$_\mathrm{b}$) to the APD is provided by a step-up converter. An avalanche pulse is divided into two (top and bottom) by 110 $\Omega$ sense resistors (R$_\mathrm{1}$,R$_\mathrm{2}$) and quenched by a 560 k$\Omega$ resistor (R$_\mathrm{q}$). The pulse heights V$_{1}$(V$_{2}$) are compared to V$_{\mathrm{ref1}}$(V$_\mathrm{ref2}$) at the constant level discriminators (CLD). Pulse stretchers (PS) convert CLD output to 150 ns duration for accumulation at two counters. The micro-controller adjusts the value of V$_\mathrm{b}$ via a Digital-to-Analog Converter (DAC) keeping the ratio of bottom-to-top counts within a pre-defined window resulting in a fixed excess voltage (V$_\mathrm{E}$).  } \label{fig:wincom} \end{figure}
 
To overcome cubesat constraints, we have developed a low power circuit for operating Geiger mode APDs with a novel mechanism for dynamic adjustment of V$_{\mathrm{b}}$ values ensuring that a fixed excess voltage (and therefore detection efficiency) is applied to an APD despite a drifting background temperature. This circuit is based on the window comparator mechanism (see Fig.~\ref{fig:wincom}) and requires only a few additional components on top of the conventional passive quenching circuit.
The window comparator mechanism works by dividing the avalanche pulse into two (top and bottom pulses) via a resistor ladder. 
The working principle is to trace the change of the avalanche pulse amplitude to the ratio of bottom-to-top pulses and to adjust the bias voltage accordingly such that the ratio is maintained within a given window keeping V$_{\mathrm{E}}$ constant. This ensures that avalanche pulses have a constant amplitude for different temperatures.

For example, the desired operating condition is set at V$_\mathrm{E}=10\mathrm{V}$. Each avalanche pulse is sampled at two resistors R$_{1}$ and R$_{2}$.  The corresponding height of the pulses V$_{1}$ and V$_2$ are compared against reference levels V$_{\mathrm{ref1}}$ and V$_\mathrm{ref2}$ by the constant level discriminators (CLD). The level V$_{\mathrm{ref1}}$ is selected so that pulses from R$_1$ are transmitted by the CLD for a wide range of V$_\mathrm{E}$ values, and accumulated in the top counter of the micro-controller.  V$_{\mathrm{ref2}}$ is chosen such that when V$_\mathrm{E}=10\mathrm{V}$, only 80\% of the bottom pulses are transmitted, resulting in a stable operating condition where the ratio of bottom-to-top counts is 0.8. 
The mechanism tolerance is defined by a user-selected window of accepted ratios, typically set between 0.8 and 0.85.  
In actual operation when the circuit is first switched on the ratio of bottom-to-top counts is very low and the bias voltage is increased until the desired ratio is reached.  
If the ratio is not in this window this indicates that the V$_\mathrm{E}$ values are outside our desired operating range and the bias voltage is adjusted accordingly.  
This mechanism enables the operating values of V$_\mathrm{E}$ to be controlled via a single parameter (the bottom-to-top ratio) that can be set by software.

This window comparator mechanism was tested for APDs operated over a range of temperatures.
The pulse height values and distribution was observed to be steady over a 30 degree range in temperature (see Fig.~\ref{fig:tempapd}(b)). 
A complete implementation of the window comparator mechanism for detecting photon pairs with two APDs (inclusive of step-up converter, CLD, PS and micro-controller) requires approximately 1W. This leaves 0.5W of electrical power in the cubesat for operating a pump laser in a correlated photon experiment.  This setup was utilised in a low power, lightweight demonsration in a high altitude experiment where a balloon carried Geiger mode APDs up to a maximum altitude of 37~km \cite{morong12}.  

\section{Space radiation environment}
\label{sec:radiation} Space radiation consisting mainly of electrons, protons and heavy ions is the greatest degradation factor for spacecraft electronics \cite{SROUR1988,Johnston2000}.  The vast majority of radiation flux in LEO comes from trapped particles in two radiation belts above the Earth \cite{sun97}.  Radiation flux has two major effects on Si APDs: total ionizing dose damage and accumulated displacement damage within the Si lattice caused primarily by protons.  In both cases, the effect on Si APDs is an increased rate of dark counts and a possible shift in the breakdown voltage.

                \begin{figure}[!h]
                        \includegraphics[width=\textwidth]{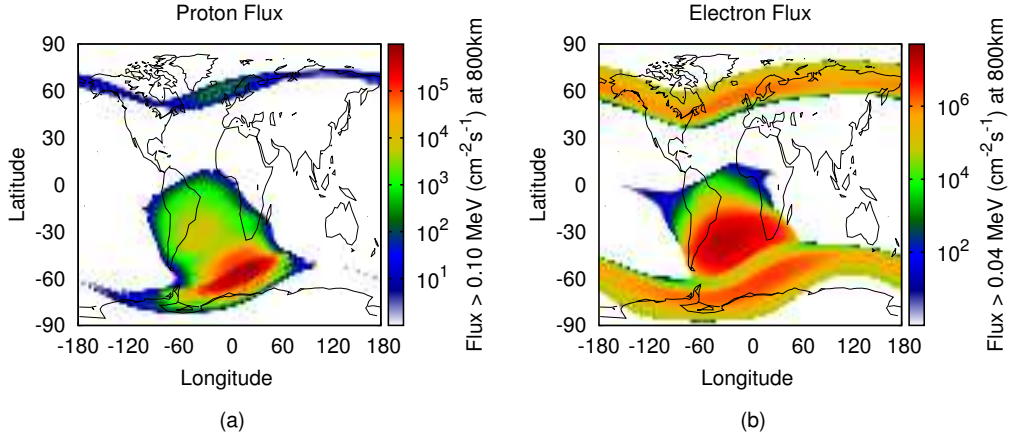}
                    \caption{Radiation flux at 800 km altitude and 98 degree inclination. The South Atlantic Anomaly (SAA) is the region of increased radiation in the southern hemisphere. (a) Proton flux via the AP-8 Max model. (b) Electron flux via the AE-8 Max model. (Color online).}
                    \label{fig:trapflux}
                \end{figure}

The Space Environment Information System (SPENVIS) \cite{spenvis} is used to analyse in-orbit radiation flux for LEO satellites.  Data from SPENVIS reveals an important feature in LEO radiation known as the South Atlantic Anomaly (SAA), a region where an inner radiation belt extends down to an altitude of 200~km due to the mismatch between the Earth's magnetic and rotational axes (see Fig.~\ref{fig:trapflux}).  Accumulated radiation effects and the resultant electronics lifetime is a function of both altitude and orbit trajectory.  

\subsection{Ionizing damage test and results}
\label{sec:ionizing}
                \begin{figure}[!h] \centering \includegraphics[width=0.85\linewidth,clip]{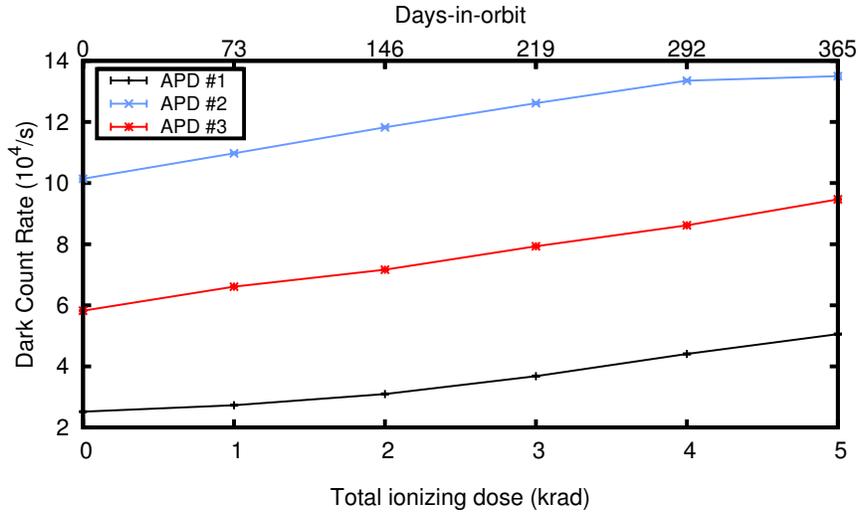} \caption{Device dark count rate after increasing ionizing dose. The number of days-in-orbit necessary to acquire the radiation dose is indicated. Three devices from the same batch were used in this test. (Color online).} \label{fig:APDrad} \end{figure}

Energetic particles traversing a solid can ionize the material in its surrounding path, generating electron-hole pairs. Under an applied electric field, the more mobile electrons are swept out immediately, and the holes are left behind. 
This causes charge accumulation which changes the electrical properties of the devices \cite{SROUR1988,Johnston2000}.
From SPENVIS data, a satellite payload at 800~km with a 98 degree inclination behind 1.85~mm of Al shielding experiences a total ionizing dose of approximately 5~krad per year.
We conducted ionizing damage testing with $\gamma$-radiation in a $^{60}$Co chamber (dose rate of 8.8 krad/hr) at the Centre for Ion Beam Application (CIBA), National University of Singapore. Three Si APDs were irradiated with $\gamma$-radiation in steps of 1~krad up to 5~krad.

In this test series, we investigated the effects of ionizing radiation on breakdown voltage and the dark count rate. 
We applied the step-stress approach whereby measurements were made after each predetermined dosage was reached using the window comparator mechanism described in the previous section.
All the APDs were stored and operated at a temperature of 22$\pm2^{\circ}$C. 
From Fig.~\ref{fig:APDrad} all tested APDs showed an increase in the dark count rate and the total increase was approximately the same for all devices despite having extensive differences in the starting rate.
All devices were able to continue to operate after irradiation up to 5krad of total dosage.  
No significant change in the breakdown voltage was observed.
Repeated attempts at annealing (55 $^\circ$C for 24 hours following \cite{sun97}) had no effect on the dark count rate revealing that ionizing damage is immune to recovery by heat.
We conclude that the total ionizing dose over one year is insufficient to halt the operation of Si APDs.

\subsection{Displacement damage test and results}
\label{sec:displacement}

Displacement damage is the result of energetic particles displacing atoms from their lattice structure \cite{SROUR1988} giving rise to intermediate energy levels in the bandgap.  For APDs this will lead to an increased rate of dark counts and may shift the breakdown voltage.  Displacement damage is stronger for lower-energy particles that can become embedded in the lattice.  The proton induced damage can be estimated by their Non-Ionizing Energy Loss [MeV $\mathrm{cm}^2$/g] (NIEL).
The displacement damage dose for a unit Si sample is the product of the fluence and the NIEL at each proton energy \cite{Raymond1987,Akkerman2001,Pease1988,dale92}.

                \begin{figure}[!h] \includegraphics[width=\textwidth]{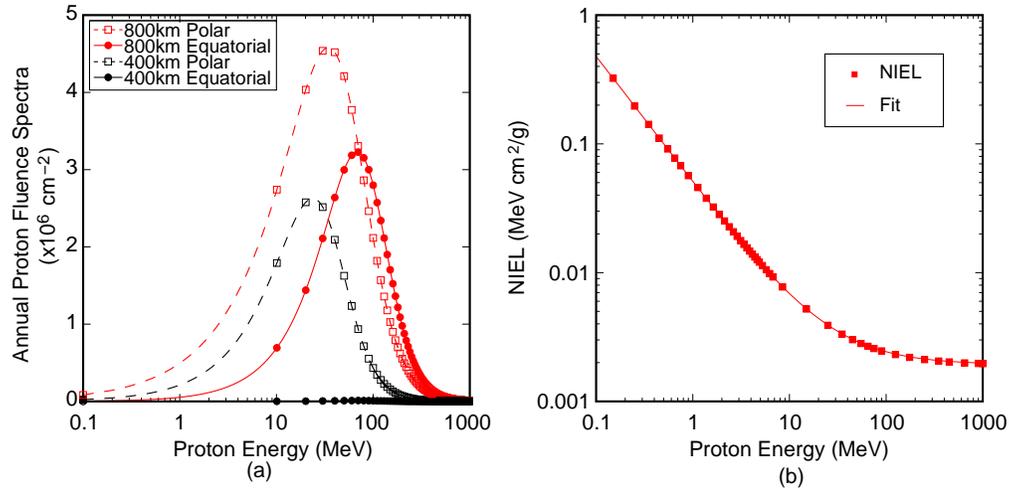} \caption{(a) Annual proton fluence spectra for various orbits with 1.85~mm of Al shielding from SPENVIS (100 keV resolution). The maximum fluence for the 400 km equatorial orbit is on the order of $10^4$. (b) Proton Non-Ionizing Energy Loss (NIEL) for silicon averaged from \cite{Summers1993,Jun2003a}. (Color online).} \label{fig:protonspectra} \end{figure}
The proton flux experienced by spacecraft depends on their altitude and inclination.  
In Fig.~\ref{fig:protonspectra}(a), we present SPENVIS data showing the spectrum of proton energy at different orbits, behind 1.85~mm of Al shielding.  We define polar and equatorial orbits as having inclinations of 98 and 20 degrees respectively.  The estimated NIEL at each proton energy is presented in Fig.~\ref{fig:protonspectra}(b).  These plots indicate that the flux of highly damaging low-energy protons (below 100 MeV) is greatly reduced in an equatorial orbit. 
The proton energy spectra and the NIEL data enable an estimate of the displacement damage dose as time progresses during a LEO mission.
Devices exposed to different doses can have their damage effects mapped to an equivalent number of days-in-orbit for different trajectories.

To estimate the effect of proton displacement damage on Si APDs, 12 devices were exposed to proton beams generated with the cyclotron at the Crocker Nuclear Laboratory (CNL) in the University of California at Davis. 
The APDs were divided into three groups a, b and c (four devices per group) and irradiated with proton beams of 5, 25 and 50 MeV respectively (see Table~\ref{tab:protonapduse}).  
The devices were calibrated at room temperature using the window comparator mechanism before being shipped to CNL for proton irradiation, after which the devices were shipped back to the lab for measurements on the post-radiation dark count rate and breakdown voltage.  
The devices were then annnealed at 55$^{\circ}$C for 24 hours (following \cite{sun97}), after which the dark count rate and breakdown voltage were re-measured. 
Finally the dark count rate was measured when all devices were cooled to -20$^{\circ}$C.

\begin{table}[!h] \centering 

\caption{Proton fluence experienced by test devices.  A total of 12 devices were tested in 3 groups (a,b,c). Each group was exposed to protons of a particular energy.  Within each group, individual devices (1-4) were exposed to different fluences. Devices are then tracked by group and number (a1 to c4).} 
\begin{tabular}{l|c|c|c|c}\hline \multicolumn{5}{c}{Test Fluence ($\times 10^{8}$cm$^{-2}$)}\\\hline \backslashbox{Group}{No.} & 1 & 2 & 3 & 4 \\  \hline a (5 MeV) & 1.86 & 3.74 & 5.61 & 7.48 \\ \hline b (25 MeV) & 1.73 & 3.46 & 5.19 & 6.92 \\ \hline c (50 MeV) &1.47 & 2.94 & 4.41 & 5.88 \\ \hline\hline \end{tabular} 
\label{tab:protonapduse} \end{table}
                
\begin{figure}[!h] \centering \includegraphics[width=\linewidth,clip]{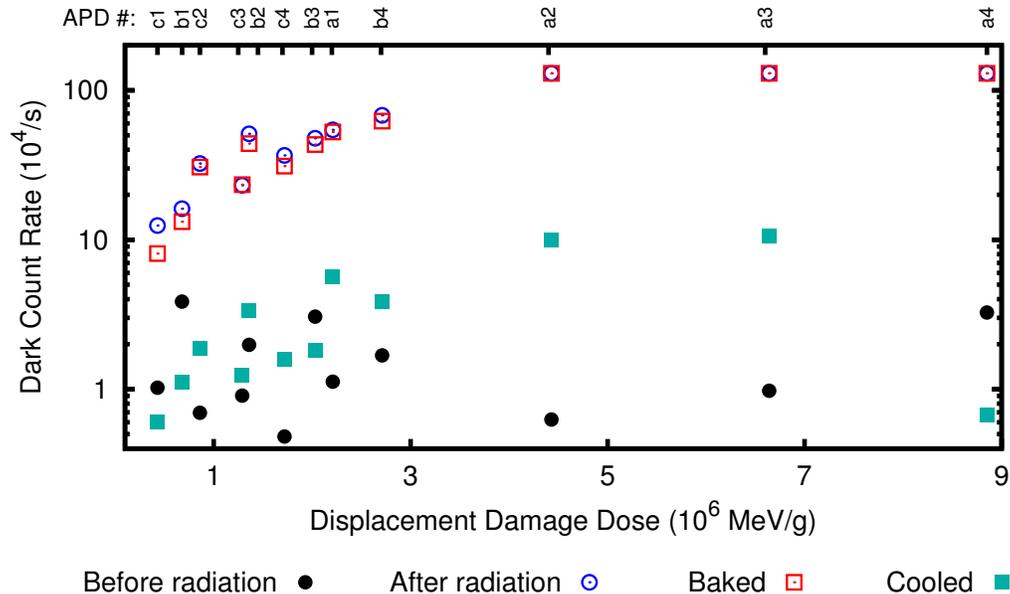} \caption{Post-irradiation dark count rate against displacement damage dose (See Table~1). Devices were operated between 22-25$^\circ$C unless cooled (-20$^\circ$C). Above $4\times 10^6$ MeV/g uncooled devices registered dark counts over $1\times 10^6$ cps, saturating the passive quenched window comparator. (Color online).} \label{fig:averageHV} \end{figure}

The observed dark count before and after irradiation for all devices are presented in Fig.~\ref{fig:averageHV}.
All 12 APDs continued to function after irradiation. 
There were no significant changes observed in the breakdown voltage.
The APD dark counts increased by one to two orders of magnitude depending on the accumulated damage.
Annealing caused partial recovery confirming that SAP500 devices followed the trend observed in \cite{sun97,sun04} but becomes ineffective after a dose of $3\times 10^6$ MeV/g.
After a dose greater than $4\times 10^6$ MeV/g, devices operated between 22-25$^\circ$C have a dark count rate sufficient to saturate the window comparator mechanism.
Cooling the devices to -20$^\circ$C, however, is extremely effective and even after the maximum dose of $9\times 10^6$~MeV/g the dark counts remain in the range of 100 kcps.

\begin{figure}[!h] \centering \includegraphics[width=0.85\linewidth,clip]{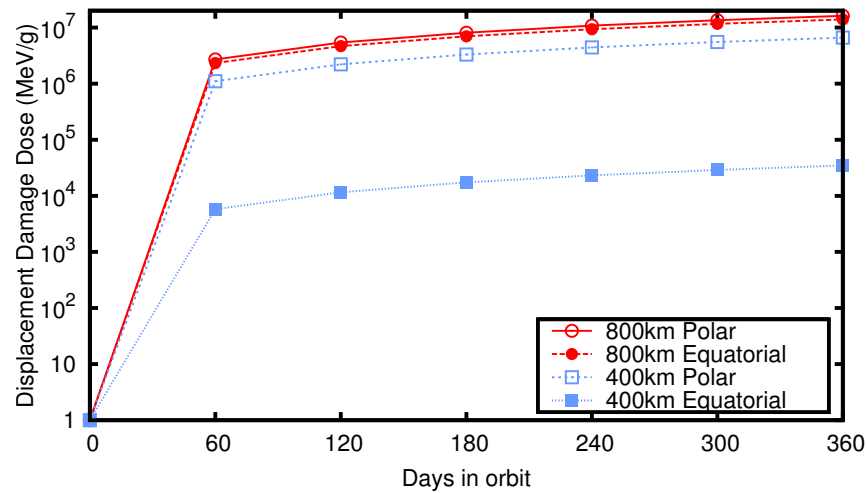} \caption{Displacement damage dose for 4 different orbits.  A low altitude equatorial orbit has a reduced radiation environment. (Color online).} \label{fig:disp} \end{figure}

It is interesting to compare the rate of displacement damage dose buildup for different orbits.  
We present this data in Fig.~\ref{fig:disp}.
In particular the radiation flux for a 400 km equatorial orbit is sufficiently low that essentially we expect the APD devices to operate without need for annealing or cooling within the effective lifetime of the spacecraft (1-2 years).  This suggests that in low resource spacecraft such as cubesats, choice of orbit will have a large infuence on the lifetime of photon counting experiments.

\section{Conclusion and outlook}
\label{sec:conclusion}
We have introduced a novel low power approach to the normal passive quenching circuit for APDs enabling them to be used in low resource cubesats where in-orbit temperature fluctuations cannot be effectively controlled.  
This technique does not rely on sensing temperature but compares the avalanche pulse amplitude via a window comparator mechanism keeping the excess voltage V$_\mathrm{E}$ (and therefore detection efficiency) of the device constant.  
Such a power efficient passive quenching circuit will be useful wherever APDs need to be operated in a low resource environment.
The window comparator can be adapted to active quenching circuits. 

We have also presented results for Si APD operation using the window comparator mechanism after the devices have been exposed to radiation damage.
The conclusions discussed below must be balanced against the small sample size used, and the possibility that different device batches may have significant differences in their response to radiation.  
Estimates of device in-orbit lifetime must be read against the reality that radiation flux can vary significantly.
Nevertheless certain trends are evident.

The tests reveal that APD breakdown voltages are not affected by radiation damage.  
The dark count rate increases for all radiation damage but over the same in-orbit duration ionizing dose has a weaker effect than proton induced displacement damage.
Annealing enables only partial recovery of the displacement damage.
Cooling the devices to -20$^{\circ}$C during operation remains the most effective method to mitigate displacement damage effects.  
In harsh orbits, a possible scenario is periodic heating of the APDs to 55$^{\circ}$C followed by cooling during operation to prolong their lifetime in orbit. However, this depends on the availability of resources on the spacecraft.  One future direction is to investigate the suitability of Si APDs that have in-built thermo-electric coolers for the active area only, as well as to investigate Si APDs with smaller active areas and thin junctions resulting in reduced accumulated radiation damage. Another area of study is the long-term effect of radiation induced damage on the detection efficiency of Si APDs. 

The radiation flux experienced by spacecraft electronics is influenced by orbital altitude and inclination.
The lifetime of Si APDs in orbit can be greatly improved by launching them into orbits of smaller inclination and/or lower altitude, and scientific experiments involving uncooled Geiger mode Si APDs can still be operated for long durations on cubesats if appropriate orbits are selected.
Analysis of the LEO radiation environment suggests uncooled Si APDs with thin shielding in a 400 km equatorial orbit can operate for several years before dark counts saturate the quenching circuit.

\section*{Acknowledgments}
The authors wish to thank S. Sachidananda for preparatory work on the window comparator \cite{subash11} and Prof. T. Osipowicz for access to a $^{60}$Co $\gamma$-chamber. Y.C. Tan and C. Cheng are supported through a DSO-NUS(CQT) Grant. 

\end{document}